\documentclass[preprint,prb,showpacs,showkeys]{revtex4}%
\usepackage{amsfonts}
\usepackage{amsmath}
\usepackage{amssymb}
\usepackage{graphicx}%
\begin{document}
\title[oscillating universe \ ]{Statistical mechanical theory of an oscillating isolated system. \\The relaxation to equilibrium}
\author{A. P\'{e}rez-Madrid}
\affiliation{Departament de F\'{\i}sica Fonamental , Facultat de F\'{\i}sica, Universitat
de Barcelona. Diagonal 647, 08028 Barcelona, Spain}

\begin{abstract}
In this contribution we show that a suitably defined nonequilibrium entropy of
an N-body isolated system is not a constant of the motion in general and its
variation is bounded, the bounds determined by the thermodynamic entropy
,\textit{i.e. , }the equilibrium entropy. We define the nonequilibrium entropy
as a convex functional of the set of $n$-particle reduced distribution
functions ($n\leq N$) generalizing the Gibbs fine-grained entropy formula.
Additionally, as a consequence of our microscopic analysis we find that this
nonequilibrium entropy behaves as a free entropic oscillator. In the approach
to the equilibrium regime we find relaxation equations of the Fokker-Planck
type, particularly for the one-particle distribution function.

\end{abstract}
\maketitle
\tableofcontents

\section{\bigskip Introduction}

It is a widely recognized fact that a general mathematical theoretical proof
of the second law is still lacking. As stated in Ref. \cite{cohen} and quoted
here just for illustration sake "To the best of our knowledge no theoretical
mathematical derivation of the second law has been given up until now; instead
it has been based on Kelvin's or Clausius's principles of the impossibility of
perpetual motion of the second kind \cite{kampen}, which are based on
experiment\cite{crooks}". This lack of definitive theoretical proof has lead
to reports on the violation of the second law \cite{evans} or tests over its
validity in some particular cases\cite{abe}, \cite{ford}.

The first significant contribution to the interpretation of the second law of
Thermodynamics and the explanation of irreversibility goes back to Boltzmann.
Nevertheless, it is known that Boltzmann's contribution was criticized by
arguing that this contradicts the predictions based on the microscopic
equations of motion. Later, Gibbs and P. Ehrenfest \& T. Ehrenfest worked on
this problem by introducing coarse-graining. However, those coarse-graining
analyses require the introduction of a \textit{priori }equal probability
principles, which are hard to justify on physical grounds as was criticized by
Einstein\cite{cohen2}.

In this scenario, our contention is to discern the connection between the
microscopic description of an isolated N-body system given through the
classical Hamiltonian dynamics and the description at the macroscopic level
expressed by \ the second law.

It is known that for thermodynamic equilibrium the entropy can \ be given by
the Gibbs formula
\begin{equation}
S_{N}=-k_{B}\text{Tr}\left(  F\ln F\right)  \text{ \ \ ,} \label{Gibbs}%
\end{equation}
where $k_{B}$ is the Boltzmann constant and $F$ the full phase-space
distribution function which we assume normalized to unity, \textit{i.e.
}Tr$\left(  F\right)  =1$. However, this expression is not adequate for
representing the entropy of nonequilibrium isolated systems for which no bath
is present \cite{Bogoliubov-2}. The reason is that although in the case of a
time-dependent distribution function out of equilibrium, the entropy $S$ given
through (\ref{Gibbs}) remains constant. This is not difficult to show given
that $F$ evolves according to the Liouville equation
\begin{equation}
\frac{\partial}{\partial t}F\mathbf{=}\left[  H,F\right]  _{p}\text{ \ ,}
\label{liouville}%
\end{equation}
where $\left[  ..,..\right]  _{P}$ is the Poisson bracket. In fact, by using
Eq. (\ref{liouville}) the rate of change of the entropy (\ref{Gibbs}) is
\begin{align}
\frac{\partial S_{N}}{\partial t}  &  =\text{Tr}\left(  \frac{\partial
F}{\partial t}\ln F\right)  =\nonumber\\
\text{Tr}\left(  \left[  H,F\right]  _{p}\ln F\right)   &  =-\text{Tr}\left(
F\left[  H,\ln F\right]  _{p}\right)  =-\text{Tr}\left(  \left[  H,F\right]
_{p}\right)  =0\text{ \ .} \label{constant_entropy}%
\end{align}

Therefore here we will generalize the Gibbs's statistics to account for the
entropy variations in nonequilibrium systems. This constitutes an application
of \ our previous results\cite{agusti}. Our starting point is the description
of the state of an isolated N-body system in terms of the set of $n$-particle
reduced distribution functions in the framework of the BBGKY
[Bogolyubov-Born-Green-Kirkwood-Yvon] description \cite{bogoliubov}. Unlike
equilibrium, an overall picture in terms of the full phase-space distribution
function does not contain the amount of detail necessary to describe a
nonequilibrium system. Nonequilibrium systems manifest a random clusterization
which makes their distribution in the phase space unstable, thus there is a
continuous process of creation of $n$-particle clusters at the expense of the
annihilation of $p$-particle clusters with $n\neq p$. This fact is taken into
account in the BBGKY hierarchy making this an appropriate framework for the
description of nonequilibrium systems. In this context, since the collisions
become explicit through the collision term in the equations of motion, the
$n$-particle reduced distribution functions are not constant of the motion,
therefore a way of defining the entropy to embody the approach to equilibrium
might be expressed in terms of \ this set of reduced distributions. This is
what we do here: we propose a functional of the set of n-particle reduced
distribution functions which generalizes the Gibbs entropy as the
nonequilibrium entropy of the isolated N-body system. We will show that this
entropy is not a constant of the motion and reaches its maximum value at equilibrium.

In the next section, we introduce the Hamiltonian dynamics of the N-body
system and obtain the generalized Liouville equation. In section 3, we define
the nonequilibrium entropy analyzing its properties. Section 4 is devoted to
computing the entropy production and to the derivation of the kinetic equation
for the one-particle reduced distribution function. In section 5 we describe
the approach to equilibrium. Finally in section 6, we emphasize our main conclusions.

\section{\bigskip Hamiltonian dynamics}

Let us consider an N-body system with a Hamiltonian containing a kinetic
energy term plus a two-particle interaction potential

\bigskip%
\begin{equation}
H=\sum_{j=1}^{N}\frac{\mathbf{p}_{j}^{2}}{2m}+\frac{1}{2}\sum_{j\neq k=1}%
^{N}\phi\left(  \left\vert \mathbf{q}_{j}-\mathbf{q}_{k}\right\vert \right)
\text{ ,} \label{hamiltonian}%
\end{equation}
with $m$ being the mass of a particle, and $\phi\left(  \left\vert
\mathbf{q}_{j}-\mathbf{q}_{k}\right\vert \right)  \equiv\phi_{j,k}$ the
interaction potential. Moreover, the equations of motion are%
\begin{equation}
\overset{\cdot}{\mathbf{q}}_{i}=\frac{\partial H}{\partial\mathbf{p}_{i}%
}\text{ \ , }\overset{\cdot}{\mathbf{p}}_{i}=-\frac{\partial H}{\partial
\mathbf{q}_{i}}\text{ \ .} \label{eq_motion}%
\end{equation}
As said in the introduction, the statistical description of the system can be
performed in terms of the full phase-space distribution function $F(x^{N},t)$,
where $x^{N}=\left\{  x_{1},...,x_{N}\right\}  $ and $x_{j}=\left(
\mathbf{q}_{j},\mathbf{p}_{j}\right)  $ or alternatively in terms of the
distribution vector\cite{balescu} $\mathbf{f}$. Both previous descriptions are
completely equivalent, however the second one is more appropriate for
nonequilibrium systems. Here,
\begin{equation}
\mathbf{f}\equiv\left\{  f_{o},f_{1}(x_{1}),f_{2}(x^{2}),.........,f_{N}%
(x^{N})\right\}  \text{ } \label{distribution}%
\end{equation}
is the set of all the $n$-particle reduced distribution functions, with
$x^{n}=\left\{  x_{1},...,x_{n}\right\}  $, $n=0,......,N$ and where the
$n$-particle reduced distribution functions%

\begin{equation}
f_{n}=\int F(x^{N},t)\text{ }dx_{n+1}...dx_{N}\text{,} \label{reduced_dis}%
\end{equation}
are obtained by integrating over the $N-n$ particles with $f_{o}=1$. The
dynamics of the reduced distribution vector follows from the Liouville
equation (\ref{liouville}) by integration according to Eq. (\ref{reduced_dis}%
), thus one obtains\cite{balescu}$^{,}$\cite{kreuzer}$^{,}$\cite{woods}%
\begin{equation}
\frac{\partial}{\partial t}f_{n}=\left[  H_{n},f_{n}\right]  _{p}+\left(
N-n\right)  \sum_{j=1}^{n}\int\mathbf{F}_{j,n+1}\frac{\partial}{\partial
\mathbf{p}_{j}}f_{n+1}dx_{n+1}\text{ ,} \label{s-part_evol}%
\end{equation}
where $\mathbf{F}_{j,n+1}=-\mathbf{\nabla}_{j,n+1}\phi_{j,n+1}$and
\begin{equation}
H_{n}=\sum_{i=1}^{n}\frac{\mathbf{p}_{i}^{2}}{2m}+\frac{1}{2}\sum_{i\neq
k=1}^{n}\phi_{i,k}\text{ } \label{s-par_hamil}%
\end{equation}
is the $n$-particle Hamiltonian.

In a compact way and in the language of Hilbert spaces we can write Eq.
(\ref{s-part_evol})\cite{balescu}$^{,}$\cite{agusti}%
\begin{equation}
i\frac{\partial}{\partial t}\mathbf{f(}t\mathbf{)=}\mathcal{L}\mathbf{f(}%
t\mathbf{)} \label{gen_lio}%
\end{equation}
constituting the generalized Liouville equation which succinctly expresses the
BBGKY\ hierarchy of equations. Here, $\mathcal{L}$ is the generalized
Liouvillian, a nonHermitian operator whose diagonal part $\mathcal{PL}$ is
defined through\cite{agusti2}$^{,}$\cite{balescu}%
\begin{equation}
\langle n\left\vert \mathcal{PL}\right\vert n^{\prime}\rangle=i\left[
H_{n},f_{n}\right]  _{P}\delta_{n^{\prime},n}\text{, }n>0\text{ ,}
\label{diago}%
\end{equation}
where $\left\vert n\right\rangle $ represents the $n$-particle state. In
addition, the nondiagonal part $\mathcal{QL}$ is given by \cite{agusti2}$^{,}%
$\cite{balescu}%
\begin{equation}
\left\langle n\right\vert \mathcal{QL}\left\vert n^{\prime}\right\rangle
=i\left\{  \left(  N-n\right)  \sum_{j=1}^{n}\int\mathbf{F}_{j,n+1}%
\frac{\partial}{\partial\mathbf{p}_{j}}f_{n+1}dx_{n+1}\right\}  \text{ }%
\delta_{n^{\prime},n+1}\text{ , }\ n>1\text{.} \label{nodia}%
\end{equation}
Here, $\mathcal{P}$ and $\mathcal{Q}$, its complement with respect to the
identity, are projector operators. From its definition through Eq.
(\ref{diago}) one can see that $\mathcal{PL}$ is a $\left(  N+1\right)
\times$ $\left(  N+1\right)  $ diagonal block Hermitian matrix. On the other
hand, from Eq. (\ref{nodia}) it is possible to infer that $\mathcal{QL}$ is a
nonHermitian $\left(  N+1\right)  \times$ $\left(  N+1\right)  $ diagonal
block matrix with nonzero elements only along the diagonal $\left(
n,n+1\right)  $ with $n>1$\cite{balescu}. In terms of the projectors just
introduced, Eq. (\ref{gen_lio}) can be rewritten%
\begin{equation}
i\frac{\partial}{\partial t}\mathbf{f(}t\mathbf{)}-\mathcal{PL}\mathbf{f(}%
t\mathbf{)=}\mathcal{QL}\mathbf{f(}t\mathbf{)}\text{ .} \label{gen_lio_2}%
\end{equation}
Hence, the formal solution of Eq. (\ref{gen_lio_2}) can be written as an
integral equation%
\begin{equation}
\mathbf{f(}t\mathbf{)=}\exp\left(  -i\mathcal{PL}t\right)  \mathbf{f(}%
0\mathbf{)}+\exp\left(  i\mathcal{PL}t\right)  \int_{0}^{t}d\tau\exp\left(
-i\mathcal{PL}\tau\right)  (-i\mathcal{QL)}\mathbf{f(}\tau\mathbf{)}\text{ }
\label{solution}%
\end{equation}
which can be formally solved to give\cite{balescu}
\begin{equation}
\mathbf{f(}t\mathbf{)=}\mathcal{U}\left(  t,0\right)  \mathbf{f(}%
0\mathbf{)}\text{ \ ,} \label{pertur_sol}%
\end{equation}
where the evolution operator $\mathcal{U}\left(  t,0\right)  $ is given by a
perturbative development as
\begin{gather}
\mathcal{U}\left(  t,0\right)  \mathbf{=}\sum_{j=0}^{\infty}\int_{0}^{t}%
dt_{1}\int_{0}^{t_{1}}dt_{2}\int_{0}^{t_{2}}dt_{3}.......\int_{0}^{t_{j-1}%
}dt_{j}\times\nonumber\\
\mathbf{V}\left(  t,t_{1}\right)  ......\mathbf{V}\left(  t_{j-1}%
,t_{j}\right)  \exp\left(  -i\mathcal{PL}t_{j}\right)  \text{ \ .}
\label{pert_oper}%
\end{gather}
Here, $\mathbf{V}\left(  t_{j-1},t_{j}\right)  =\exp\left[  i\mathcal{PL}%
\left(  t_{j-1}-t_{j}\right)  \right]  (-i\mathcal{QL)}$ are nonHermitian
propagators, $t_{j}<t_{j-1}<.....<t_{1}<t_{0}=t$, and the integration proceeds
from right to left. Differentiating Eq. (\ref{pert_oper}) one find
\begin{equation}
\frac{\partial}{\partial t}\mathcal{U}\left(  t,0\right)  =-i\mathcal{LU}%
\left(  t,0\right)  \text{ ,} \label{evol_1}%
\end{equation}
the evolution equation for $\mathcal{U}\left(  t,0\right)  $. If now, we make
a time-translation, and change the origin of the time scale so that the time
series begin at time $t$%
\begin{gather}
\mathcal{U}\left(  0,-t\right)  \mathbf{=}\sum_{j=0}^{\infty}\int_{-t}%
^{0}dt_{1}^{\prime}\int_{-t}^{t_{1}^{\prime}}dt_{2}^{\prime}\int_{-t}%
^{t_{2}^{\prime}}dt_{3}^{\prime}.......\int_{-t}^{t_{j}^{\prime}}%
dt_{j}^{\prime}\times\nonumber\\
\mathbf{V}\left(  0,t_{1}^{\prime}\right)  ......\mathbf{V}\left(
t_{j-1}^{\prime},t_{j}^{\prime}\right)  \exp\left[  -i\mathcal{PL}\left(
t_{j}^{\prime}+t\right)  \right]  \text{ \ ,} \label{pert_oper_2}%
\end{gather}
with $t_{l}^{\prime}=t_{l}-t$, $(1\leq l\leq j)$ and under time reversal%
\begin{gather}
\mathcal{U}\left(  0,t\right)  \mathbf{=}\sum_{j=0}^{\infty}\int_{t}^{0}%
dt_{1}\int_{t}^{t_{1}}dt_{2}\int_{t}^{t_{2}}dt_{3}.......\int_{t}^{t_{j-1}%
}dt_{j}\times\nonumber\\
\mathbf{V}\left(  0,t_{1}\right)  ......\mathbf{V}\left(  t_{j-1}%
,t_{j}\right)  \exp\left[  -i\mathcal{PL}\left(  t_{j}-t\right)  \right]
\text{ \ .} \label{pert_oper_3}%
\end{gather}
Interchanging the integration limits and the integrals we obtain after
relabeling the dummy integration variables\cite{evans2}
\begin{gather}
\mathcal{U}\left(  0,t\right)  \mathbf{=}\sum_{j=0}^{\infty}\left(  -1\right)
^{j}\int_{0}^{t}dt_{1}\int_{0}^{t_{1}}dt_{2}\int_{0}^{t_{2}}dt_{3}%
.......\int_{0}^{t_{j-1}}dt_{j}\times\nonumber\\
\exp\left(  -i\mathcal{PL}t_{1}\right)  \mathbf{U}\left(  t_{j},t_{j-1}%
\right)  ......\mathbf{U}\left(  t_{1},t\right)  \text{ \ ,}
\label{pert_oper_4}%
\end{gather}
where now, $\mathbf{U}\left(  t_{j},t_{j-1}\right)  =$ $(-i\mathcal{QL)}%
\exp\left[  -i\mathcal{PL}\left(  t_{j}-t_{j-1}\right)  \right]  $. In
addition, by differentiating Eq. (\ref{pert_oper_4}) one gets the evolution
equation for $\mathcal{U}\left(  0,t\right)  $
\begin{equation}
\frac{\partial}{\partial t}\mathcal{U}\left(  0,t\right)  =\mathcal{U}\left(
0,t\right)  i\mathcal{L}\text{ \ .} \label{evol_2}%
\end{equation}

The propagator $\mathcal{U}\left(  0,t\right)  $ given through Eq.
(\ref{pert_oper_3}) propagates backwards in time from $t$ to $0$, hence this
must coincide with the inverse $\mathcal{U}\left(  t,0\right)  ^{-1}$ of
\ $\mathcal{U}\left(  t,0\right)  $ so that
\begin{equation}
\mathcal{U}\left(  t,0\right)  ^{-1}\mathbf{f(}t\mathbf{)}\text{ }%
=\text{\ }\mathcal{U}\left(  0,t\right)  \mathbf{f(}t\mathbf{)=f(}%
0\mathbf{)}\text{ .} \label{pertur_sol_2}%
\end{equation}
It can be verified that $\mathcal{U}\left(  0,t\right)  $ is the inverse of
$\ \mathcal{U}\left(  t,0\right)  $\cite{evans2}. To begin, $\mathcal{U}%
\left(  0,t\right)  \mathcal{U}\left(  t,0\right)  =1$ for $t=0$. Now by
differentiating and taking into account Eqs. (\ref{evol_1}) and (\ref{evol_2})
we reach
\begin{equation}
\frac{\partial}{\partial t}\mathcal{U}\left(  0,t\right)  \mathcal{U}\left(
t,0\right)  =0\text{ ,} \label{invers_cond}%
\end{equation}
so $\mathcal{U}\left(  0,t\right)  \mathcal{U}\left(  t,0\right)  =1$ for all
$t$.

Since $\mathcal{PL}$ is Hermitian, all its eigenvalues are real\cite{tolman},
which means that $\mathbf{f(}t\mathbf{)}$ as given through Eqs.
(\ref{pertur_sol}) and (\ref{pert_oper}) will have an oscillatory behavior.

\section{Nonequilibrium entropy}

Here as the nonequilibrium entropy for the N-body system we propose
\cite{agusti}$^{,}$\cite{agusti2}%
\begin{gather}
S=-k_{B}\text{Tr}\left\{  \mathbf{f}\ln\mathbf{f}_{eq}^{-1}\mathbf{f}\right\}
+S_{eq}\nonumber\\
=-k_{B}\sum_{n=1}^{N}\int f_{n}\ln\frac{f_{n}}{f_{eq,n}}\;dx_{1}%
.....dx_{n}\text{ }+S_{eq}\text{ \ ,} \label{noneq_entropy}%
\end{gather}
a convex functional of the distribution vector which generalizes the Gibbs
formula . In Eq. (\ref{noneq_entropy}), $S_{eq}$ is the thermodynamic entropy
(\textit{i.e. }the equilibrium entropy) and $\mathbf{f}_{eq}$ is the
equilibrium distribution vector satisfying $\mathcal{L}\mathbf{f}_{eq}=0$, the
Yvon-Born-Green (YBG) equilibrium hierarchy\cite{balescu}. Therefore,
$\mathbf{f}_{eq}$ is an eigenfunction of $\mathcal{L}$ with eigenvalue $0$.
Moreover,
\begin{equation}
\delta S=-k_{B}\text{Tr}\left\{  \delta\mathbf{f}\ln\left(  \mathbf{f}%
_{eq}^{-1}\mathbf{f}\right)  \right\}  \label{first_var}%
\end{equation}
is zero at equilibrium and
\begin{equation}
\delta^{2}S=-\frac{1}{2}k_{B}\text{Tr}\left\{  \delta\mathbf{f\;f}_{eq}%
^{-1}\delta\mathbf{f}\right\}  \label{second_var}%
\end{equation}
is a negative quantity, which shows that $S$ is maximum at equilibrium where
its value is $S_{eq}$.

Note that the BBGKY scenario describes an interacting mixture of fluids made
up of particle clusters in the phase space. Two such fluids differ in the size
of the clusters they contain and each fluid contributes its own entropy, the
$n$-particle entropy, to the total nonequilibrium entropy of our system.
Likewise, the interaction between different fluids leads to the creation of
\ \ $n$-particles clusters at the expense of the annihilation of
\ $p$-particle clusters with $n\neq p$.

More interestingly here the most important property of the entropy we propose
is its direction of change in a natural process. To elucidate this, we must
establish the entropy bounds, if any. Hence, let us define the $n$-particle
entropies%
\begin{equation}
S_{n}=-k_{B}\int f_{n}\ln\frac{f_{n}}{f_{eq,n}}\;dx_{1}.....dx_{n}\text{ .}
\label{n-part_entrop}%
\end{equation}
Since the full distribution function $F$ contains more information than
$f_{n}$, one might expect that $S_{n}\geq S_{N}$. This can be proved from the
convexity of the logarithmic function, $\ln x\leq x-1$ which can be
rewritten\cite{grad}%
\begin{equation}
f\ln f-f\ln g\geq f-g\text{ \ for }f\geq0,\text{ }g>0\text{ \ },
\label{convexity}%
\end{equation}
where there is strict inequality unless $f=g$. Hence, assuming that $f=F$ and
$g=f_{n}$, from Eq. (\ref{convexity}) one derives%
\begin{gather}
\int F\ln Fdx_{1}....dx_{N}\geq\int F\ln f_{n}dx_{1}....dx_{N}=\nonumber\\
\int f_{n}\ln f_{n}dx_{1}....dx_{n}\text{ .} \label{first_ineq}%
\end{gather}
Analogously, it can be proved%
\begin{equation}
\int f_{n}\ln f_{n}dx_{1}....dx_{n}\geq\int f_{n}\ln f_{eq,n}dx_{1}%
....dx_{n}\text{ ,} \label{second_ineq}%
\end{equation}
which allows us to rewrite Eq. (\ref{first_ineq})%
\begin{equation}
\int F\ln Fdx_{1}....dx_{N}\geq\int f_{n}\ln\frac{f_{n}}{f_{eq,n}}%
\;dx_{1}.....dx_{n}\text{ .} \label{third_ineq}%
\end{equation}
Therefore, from Eqs. (\ref{Gibbs}) and (\ref{n-part_entrop})%
\begin{equation}
S_{n}\geq S_{N} \label{fourth_ineq}%
\end{equation}
and consequently, from Eqs. (\ref{noneq_entropy}), (\ref{n-part_entrop}) and
(\ref{fourth_ineq}) one obtains%
\begin{equation}
0\geq S-S_{eq}\geq S_{N}\text{ .} \label{fifth_ineq}%
\end{equation}
In light of this, we find that $S$ is bounded%
\begin{equation}
0\geq S_{eq}\geq S\geq S_{N}+S_{eq}\text{ .} \label{bounds}%
\end{equation}
This result together with our comments at the end of the previous section
leads us to conclude that the nonequilibrium entropy $S$ behaves as a free
oscillator with an amplitude of oscillation $\bigtriangleup_{S}=-S_{N}/2$.
Hence, there is no possibility of time arrow and a question arises as to how
the equilibrium could be reached and more deeply how to re-read the second law
for isolated systems. We will try to answer these questions in the next section.

To end this section, in view of our previous conclusion we assume the
existence of a potential associated to the harmonic entropic oscillator
\begin{equation}
\Phi(S)=\frac{1}{2}K_{eff}(t)\left(  S-S^{\ast}\right)  ^{2}\text{ ,}
\label{potential}%
\end{equation}
with $S^{\ast}=S_{eq}+\frac{S_{N}}{2}$. Therefore, the potential bounds
satisfy%
\begin{equation}
\Phi(S_{eq})=\Phi(S_{N}+S_{eq})=\frac{1}{2}K_{eff}(t)\left(  \frac{S_{N}}%
{2}\right)  ^{2} \label{poten_bounds}%
\end{equation}
and the effective elastic constant is given through%
\begin{equation}
K_{eff}(t)=\left\vert \frac{1}{S}\frac{\partial^{2}}{\partial t^{2}%
}S\right\vert \text{ .} \label{elastic_cons}%
\end{equation}
Consequently,
\begin{equation}
S-S^{\ast}=\bigtriangleup_{S}\sin\left(  \sqrt{K_{eff}(t)}t+\varphi\right)
\text{ ,} \label{oscillator}%
\end{equation}
where $\varphi$ stands for the initial conditions. This system has a first
integral of the motion its 'energy' given by\cite{gentile}%
\begin{equation}
\mathcal{H}\left(  S(t),\dot{S}(t),t\right)  =\frac{1}{2}\left(  \frac{\dot
{S}(t)}{\sqrt{K_{eff}(t)}}\right)  ^{2}+\Phi(S(t)) \label{energy}%
\end{equation}
which is constant. Moreover, the period of the oscillations $\tau$ satisfies
\begin{equation}
2\pi=\tau\sqrt{K_{eff}(\tau)}\text{ } \label{period}%
\end{equation}
and should be coherent with the recurrence period of the Poincare cycles.

\section{Entropy production. The law of increase of entropy}

The rate of change of the nonequilibrium entropy or entropy production is
obtained by taking the time derivative of Eq. (\ref{noneq_entropy}), giving
\begin{equation}
\frac{\partial S}{\partial t}=-k_{B}\text{Tr}\left\{  \frac{\partial
\mathbf{f}}{\partial t}\ln\left(  \mathbf{f}_{eq}^{-1}\mathbf{f}\right)
\right\}  =ik_{B}\text{Tr}\left\{  \mathcal{L}\mathbf{f}\ln\left(
\mathbf{f}_{eq}^{-1}\mathbf{f}\right)  \right\}  \text{ .} \label{entr_pro_1}%
\end{equation}
In a more explicit way, after using Eqs. (\ref{s-part_evol}) (\ref{gen_lio}%
)-(\ref{nodia}), Eq. (\ref{entr_pro_1}) can be rewritten%
\begin{equation}
\frac{\partial S}{\partial t}=-\frac{1}{T}\sum_{n=1}^{N}\sum_{j=1}^{n}\int
f_{n}\mathbf{p}_{j}\left(  -k_{B}T\frac{\partial}{\partial\mathbf{q}_{j}}\ln
f_{eq,n}+\sum_{j\neq i=1}^{n}\mathbf{F}_{j,i}+\left(  N-n\right)
\mathcal{F}_{j}\right)  dx^{n} \label{entr_pro_2}%
\end{equation}
where $\mathcal{F}_{j}(x^{n})$ is defined through $f_{n}(x^{n})$
$\mathcal{F}_{j}(x^{n})=\int\mathbf{F}_{j,n+1}f_{n+1}dx_{n+1}$, and $T$ is the
kinetic temperature taking into account that the dependence of $f_{eq,n}$ in
the velocities is given through a local Maxwellian. The entropy production
given in Eq. (\ref{entr_pro_2}) vanishes at equilibrium and in any other case
it should not be necessarily zero. In addition, because $\mathbf{p}_{j}$ is
arbitrary%
\begin{equation}
\sum_{j\neq i=1}^{n}\mathbf{F}_{j,i}+\left(  N-n\right)  \mathcal{F}_{j}%
^{eq}=k_{B}T\frac{\partial}{\partial\mathbf{q}_{j}}\ln f_{eq,n} \label{YBG}%
\end{equation}
is sufficient to satisfy the extremum condition $\delta\dot{S}/\delta
f_{n}\mid_{eq}=0$, with $\dot{S}=\partial S/\partial t$. Precisely, Eq.
(\ref{YBG}) gives rise to the YBG hierarchy\cite{agusti}$^{,}$\cite{hill}.

On the other hand, by using Eq. (\ref{YBG}), we can rewrite the entropy
production given through Eq. (\ref{entr_pro_2}) as%
\begin{equation}
\frac{\partial S}{\partial t}=-\frac{1}{T}\sum_{n=1}^{N}\sum_{j=1}^{n}\left(
N-n\right)  \int f_{n}\mathbf{p}_{j}\left(  \mathcal{F}_{j}-\mathcal{F}%
_{j}^{eq}\right)  dx^{n}\text{ \ ,} \label{entr_pro_3}%
\end{equation}
which is the starting equation to analyze the relaxation to equilibrium. To
this end, as in Nonequilibrium Thermodynamics\cite{groot}, from Eq.
(\ref{entr_pro_3}) we can establish the phenomenological relation%
\begin{equation}
f_{n}\mathbf{p}_{j}=f_{eq,n}\mathbf{p}_{j}-\sum_{i=1}^{n}\frac{\mathbf{L}%
_{j,i}}{T}\left(  \mathcal{F}_{i}-\mathcal{F}_{i}^{eq}\right)  \text{ ,}
\label{pheno}%
\end{equation}
\bigskip where $\mathbf{L}_{j,i}$ is a phenomenological matrix which in
general might depend on the nonequilibrium thermodynamic force $\left(
\mathcal{F}_{i}-\mathcal{F}_{i}^{eq}\right)  $. In terms of the mobility
$\mathbf{M}_{j,i}=\mathbf{L}_{j,i}/Tf_{n}$ we can rewrite Eq. (\ref{pheno}) as%
\begin{equation}
\mathbf{J}_{j}=-\sum_{i=1}^{n}f_{n}\mathbf{M}_{j,i}\left(  \mathcal{F}%
_{i}-\mathcal{F}_{i}^{eq}\right)  \text{ ,} \label{pheno_2}%
\end{equation}
where we have defined the current $\mathbf{J}_{j}=f_{n}\mathbf{p}_{j}%
-f_{eq,n}\mathbf{p}_{j}$, so $\mathbf{J}_{j}$ and $\left(  \mathcal{F}%
_{i}-\mathcal{F}_{i}^{eq}\right)  $ constitute a pair of conjugated current
and thermodynamic force, respectively .

Now we are in position to introduce the crucial point that might explain the
approach to equilibrium and the link with the macroscopic irreversibility.
Thus, near equilibrium, which coincides with the extremum position of the
entropic oscillator related to the potential given through Eq.
(\ref{potential}), $\mathbf{M}_{j,i}\longrightarrow\mathbf{M}_{j,i}^{eq}$
which in this particular case should be a nonnegative constant matrix.
Therefore, in this case%
\begin{equation}
\frac{\partial S}{\partial t}=\frac{1}{T}\sum_{n=1}^{N}\sum_{j=1}^{n}\left(
N-n\right)  \int f_{n}\left(  \mathcal{F}_{j}-\mathcal{F}_{j}^{eq}\right)
\mathbf{M}_{j,i}^{eq}\left(  \mathcal{F}_{i}-\mathcal{F}_{i}^{eq}\right)
dx^{n}\geq0\text{ ,} \label{second_law}%
\end{equation}
constituting the law of increase of entropy.

\section{Relaxation equations}

In this section we will analyze the relaxation to equilibrium by deriving the
relaxation equation for the one particle reduced distribution function. To
obtain such an equation we introduce the inverse mobility matrix $\zeta
_{j,i}^{eq}$ ($\sum_{i=1}^{n}\zeta_{j,i}^{eq}\mathbf{M}_{i,l}^{eq}%
=\delta_{j,l}$), the friction matrix which allows us to invert the near
equilibrium version of Eq. (\ref{pheno_2})
\begin{equation}
f_{n}\left(  \mathcal{F}_{i}-\mathcal{F}_{i}^{eq}\right)  =-\sum_{i=1}%
^{n}\zeta_{i,j}^{eq}\mathbf{J}_{j}=-\sum_{i=1}^{n}\zeta_{i,j}^{eq}%
\mathbf{p}_{j}\left(  f_{n}-f_{eq,n}\right)  \text{ .} \label{pheno_3}%
\end{equation}
At this point it will be useful to introduce the physical volume of the system
$V$ as a scale factor, thus we will redefine the reduced distribution
functions\cite{woods}.%
\begin{equation}
\hat{f}_{n}=V^{n}f_{n}\text{ .} \label{weighted_dis}%
\end{equation}
Additionally, we must also redefine the forces, writing $\mathcal{\hat{F}}%
_{i}/V$ and $\mathcal{\hat{F}}_{i}^{eq}/V$ instead of $\mathcal{F}_{i}$and
$\mathcal{F}_{i}^{eq}$. Hence, for $n=1$, we obtain from Eq.
(\ref{s-part_evol})%
\begin{equation}
\frac{\partial}{\partial t}\hat{f}_{1}+\mathbf{p}\frac{\partial}%
{\partial\mathbf{q}}\hat{f}_{1}=\left(  \frac{N-1}{V}\right)  \frac{\partial
}{\partial\mathbf{p}}\hat{f}_{1}\mathcal{\hat{F}}_{1}\text{ .}
\label{kinetic_0}%
\end{equation}
Thus, by using Eqs. (\ref{YBG}), (\ref{pheno_3}) and (\ref{kinetic_0}) we
obtain the kinetic equation for $\hat{f}_{1}\equiv f$
\begin{gather}
\frac{\partial}{\partial t}f+\mathbf{p}\frac{\partial}{\partial\mathbf{q}%
}f-k_{B}T\left(  \frac{\partial}{\partial\mathbf{q}}\ln f_{eq}\right)
\frac{\partial}{\partial\mathbf{p}}f=\nonumber\\
-\left(  \frac{N-1}{V}\right)  \zeta\frac{\partial}{\partial\mathbf{p}%
}\mathbf{p}\left(  f-f_{eq}\right)  \text{ \ ,} \label{kinetic}%
\end{gather}
where $\zeta\equiv\zeta_{2,1}^{eq}$. In the thermodynamic limit%
\begin{gather}
\frac{\partial}{\partial t}f+\mathbf{p}\frac{\partial}{\partial\mathbf{q}%
}f-k_{B}T\left(  \frac{\partial}{\partial\mathbf{q}}\ln f_{eq}\right)
\frac{\partial}{\partial\mathbf{p}}f=\nonumber\\
-\rho\zeta\frac{\partial}{\partial\mathbf{p}}\mathbf{p}\left(  f-f_{eq}%
\right)  \text{ \ ,} \label{kinetic_2}%
\end{gather}
with $\rho=N/V$ being the density. This equation constitutes a generalization
of the Bhatnagar-Gross-Krook (BGK) relaxation model\cite{agusti}.

To illustrate the approach to equilibrium let us write%
\begin{equation}
f(\mathbf{q},\mathbf{p},t)=\psi_{\mathbf{q}}(\mathbf{p},t)\phi(\mathbf{q}%
,t)\text{ .} \label{decoupling}%
\end{equation}
By introducing the factorization given through Eq. (\ref{decoupling}) into Eq.
(\ref{kinetic_2}) after integration in $\mathbf{p}$ we obtain%
\begin{equation}
\frac{\partial\phi}{\partial t}+\frac{\partial}{\partial\mathbf{q}}J_{\phi
}=0\text{ ,} \label{continuity}%
\end{equation}
where
\begin{equation}
J_{\phi}=\phi\int\psi_{\mathbf{q}}\mathbf{p}d\mathbf{p} \label{current}%
\end{equation}
is the current of the probability density $\phi(\mathbf{q},t)$ or first moment
of the density $\psi_{q}$, which satisfies the equation
\begin{equation}
\frac{\partial}{\partial t}J_{\phi}+\frac{\partial}{\partial\mathbf{q}}%
\phi\int\psi_{\mathbf{q}}\mathbf{pp}d\mathbf{p}+k_{B}T\left(  \frac{\partial
}{\partial\mathbf{q}}\ln f_{eq}\right)  \phi=\rho\zeta J_{\phi}\text{ .}
\label{current_2}%
\end{equation}
Here, for time $t\gg\left(  \rho\zeta\right)  ^{-1}$, Eq. (\ref{current_2})
leads to%
\begin{equation}
J_{\phi}=\frac{1}{\rho\zeta}\frac{\partial}{\partial\mathbf{q}}\phi\int
\psi_{\mathbf{q}}\mathbf{pp}d\mathbf{p+}\frac{k_{B}T}{\zeta}\left(
\frac{\partial}{\partial\mathbf{q}}\ln f_{eq}\right)  \phi\text{ ,}
\label{current_3}%
\end{equation}
which substituted into Eq. (\ref{continuity}) and assuming that $\psi
_{\mathbf{q}}$ is a local Maxwellian such that $\int\psi_{\mathbf{q}%
}\mathbf{pp}d\mathbf{p=}-k_{B}T$, gives%
\begin{equation}
\frac{\partial\phi}{\partial t}=-\mathbf{L}_{\mathbf{q}}\phi\text{ ,}
\label{kinetic_3}%
\end{equation}
where the lineal differential operator $\mathbf{L}_{\mathbf{q}}$ is given
through%
\begin{equation}
\mathbf{L}_{\mathbf{q}}\phi=-\frac{k_{B}T}{\rho\zeta}\frac{\partial}%
{\partial\mathbf{q}}\left(  \frac{\partial}{\partial\mathbf{q}}\phi-\phi
\frac{\partial}{\partial\mathbf{q}}\ln f_{eq}\right)  \text{ .}
\label{operator}%
\end{equation}
This equation contains a term $k_{B}T\partial/\partial\mathbf{q}\ln f_{eq}$
that plays the role of a thermal force usually introduced in polymer
dynamics\cite{bird}.

In the next section, starting in Eq. (\ref{kinetic_3}) and from the properties
of $\ \mathbf{L}_{\mathbf{q}}$ defined through Eq. (\ref{operator}) we will
study the approach to equilibrium.

\section{Approach to equilibrium}

\bigskip The differential operator $\mathbf{L}_{\mathbf{q}}$ introduced in the
previous section is a nonHermitian operator whose Hermitian conjugated is
defined through
\begin{equation}
\mathbf{L}_{\mathbf{q}}^{\dag}\varphi=-\frac{k_{B}T}{\rho\zeta}\left(
\frac{\partial}{\partial\mathbf{q}}+\frac{\partial}{\partial\mathbf{q}}\ln
f_{eq}\right)  \frac{\partial}{\partial\mathbf{q}}\varphi\text{ .}
\label{Hermitian}%
\end{equation}
Therefore, $\mathbf{L}_{\mathbf{q}}$ does not have an orthonormal base.
However, we can define right-hand and left-hand eigenfunctions through%
\begin{equation}
\mathbf{L}_{\mathbf{q}}\Omega_{p}(\mathbf{q})=\lambda_{p}\Omega_{p}%
(\mathbf{q}) \label{right}%
\end{equation}
and%
\begin{equation}
\mathbf{L}_{\mathbf{q}}^{\dag}\omega_{p}(\mathbf{q})=\lambda_{p}\omega
_{p}(\mathbf{q})\text{ ,} \label{left}%
\end{equation}
which we chose to be orthonormal, $\int d\mathbf{q}\Omega_{p}(\mathbf{q}%
)\omega_{r}(\mathbf{q})$ $=\delta_{p,r}$. As has been said in section III,
$\phi_{eq}(\mathbf{q)}$ is an eigenfunction with eigenvalue 0 of the evolution
operator $\mathbf{L}_{\mathbf{q}}$, thus it is possible to write
\begin{equation}
\Omega_{p}(\mathbf{q})=\phi_{eq}(\mathbf{q)}\omega_{p}(\mathbf{q})\text{ ,}
\label{relation}%
\end{equation}
which can be proved by direct substitution of this relation into Eq.
(\ref{right}) and taking into account Eqs. (\ref{operator}), (\ref{Hermitian})
and (\ref{left}). So, $\Omega_{0}(\mathbf{q})=\phi_{eq}(\mathbf{q)}$ and
$\omega_{0}(\mathbf{q})=1$. On the other hand the eigenvalues different from
zero are positive. In fact, by multiplying Eq. (\ref{right}) by $\omega
_{p}(\mathbf{q})$ and integrating, one obtains%
\begin{gather}
\lambda_{p}\int d\mathbf{q}\phi_{eq}\omega_{p}^{2}=\int d\mathbf{q}\omega
_{p}\mathbf{L}_{\mathbf{q}}\phi_{eq}\omega_{p}=\nonumber\\
\frac{k_{B}T}{\rho\zeta}\int d\mathbf{q}\phi_{eq}\frac{\partial}%
{\partial\mathbf{q}}\omega_{p}\frac{\partial}{\partial\mathbf{q}}\omega
_{p}\geq0\text{ ,} \label{positivity}%
\end{gather}
where the right-had side in the last equality has been obtained by using Eq.
(\ref{operator}) and integrating by parts. Also, the positivity of $\zeta$
discussed in the context of Eq. (\ref{second_law}) has been taken into account.

Now any distribution $\phi(\mathbf{q},t)$ can by expanded in terms of the
eigenfunctions%
\begin{equation}
\phi(\mathbf{q},t)=\sum_{p}\alpha_{p}(t)\omega_{p}(\mathbf{q})\phi
_{eq}(\mathbf{q)}\text{ ,} \label{expansion}%
\end{equation}
where taking into account the orthonormality condition%
\begin{equation}
\alpha_{p}(t)\mathbf{=}\int d\mathbf{q}\omega_{p}(\mathbf{q})\phi
(\mathbf{q},t)\text{ .} \label{coefficient}%
\end{equation}
From Eqs. (\ref{kinetic_3}), (\ref{operator}) and (\ref{expansion}) one
obtains the evolution equation for the coefficients of the expansion
(\ref{expansion})%
\begin{equation}
\frac{d}{d}\alpha_{p}(t)=-\lambda_{p}\alpha_{p}(t) \label{evol_coeff}%
\end{equation}
which gives%
\begin{equation}
\alpha_{p}(t)=\alpha_{p}(0)\exp(-\lambda_{p}t)\text{ .} \label{solution_3}%
\end{equation}
Since $\omega_{0}=1$ and $\phi(\mathbf{q},t)$ should be normalized,
$\alpha_{0}=1$. Therefore,%
\begin{equation}
\phi(\mathbf{q},t)=\phi_{eq}(\mathbf{q)+}\sum_{p}\alpha_{p}(0)\exp
(-\lambda_{p}t)\omega_{p}(\mathbf{q})\phi_{eq}(\mathbf{q)} \label{final}%
\end{equation}
showing that after a long period of time equilibrium is eventually reached.

\section{\bigskip Conclusions}

We find that the description of an N-body isolated system in the framework of
the BBGKY hierarchy enables us to prove that the nonequilibrium entropy is not
a constant of the motion. We emphasize that the nonequilibrium entropy should
be defined as a convex functional of the distribution vector. Our contention
is that the adequate functional is the one given in Eq. (\ref{noneq_entropy}).
Moreover, this description reconciles the reversibility of the Hamiltonian
dynamics with the approach to equilibrium.

Due to the periodic character of the solution of the microscopic equations
given through Eqs. (\ref{pertur_sol}) and (\ref{pert_oper}), we realize that
the nonequilibrium entropy \ corresponds to a dynamical system that behaves as
a free oscillator, an 'entropic oscillator' with well established bounds
determined by the equilibrium entropy which is the maximum entropy. We have
also manage to construct the Hamiltonian for this entropic oscillator. Hence,
the approach to equilibrium occurs when the entropy production is positive,
\textit{i.e.}, when the dynamical system is rising through the walls of the
elastic potential defined in Eq. (\ref{potential}). In other words, the
entropy production is positive when the balance of \ forces appearing in the
integrand of Eq. (\ref{entr_pro_2}) is opposite to velocity thus preventing
the expansion of the N-body system in the phase space. The natural extension
of our theory to study non isolated systems, \textit{i.e.} dissipative N-body
systems, would be to consider a damped entropic oscillator instead a free
oscillator. In the case of the damped entropic oscillator the system collapses
in the equilibrium state which is the attractor of the dynamics.

Performing a nonequilibrium thermodynamic analysis we are able to derive
relaxation equations of the Fokker-Planck type, particularly for the
one-particle distribution function. Finally, through an spectral analysis we
show how these equations describe the approach to equilibrium.

\end{document}